\documentclass[12pt]{article}
\usepackage{mathtext}
\usepackage{graphicx}
\usepackage[english]{babel}
\usepackage[T2A]{fontenc}
\usepackage[utf8]{inputenc}
\usepackage{amsfonts}
\usepackage{amssymb}
\usepackage{amsmath}
\usepackage{pgfplots}
\usepackage{braket}
\usepackage{tikz}
\usepackage{float}
\newcommand{\s}{\sigma}

\begin{document}
\title{Kinship and antagonism of chains in entangled biphoton controlled synthesis}

\author{Y.I. Ozhigov$^{1,2}$, I.R. Pluzhnikov$^1$\\
{\it 
1.Lomonosov Moscow State University, 
} \\
{\it faculty of computational mathematics and Cybernetics,}
\\
{\it 2. K. A. Valiev Russian Academy of Sciences, Moscow, Russia}
\\ 
}
\maketitle

\begin{abstract}
The synthesis of 3 and 4 abstract polymer chains divided into two sexes is considered, where the degree of kinship of the chains is determined by their overlap. It is shown that the use of some types of entangled bi-photon in one-way control gives a difference in the degree of kinship between the legal and nonlegal pairs that is unattainable with classical control. This example demonstrates the quantum superiority in distributed computing, coming from the violation of the Bell inequality. It may  be of interest for revealing the quantum mechanisms of synthesis of real biopolymers with directional properties.
\end{abstract}

\begin{figure}
\centering
\includegraphics[scale=0.5]{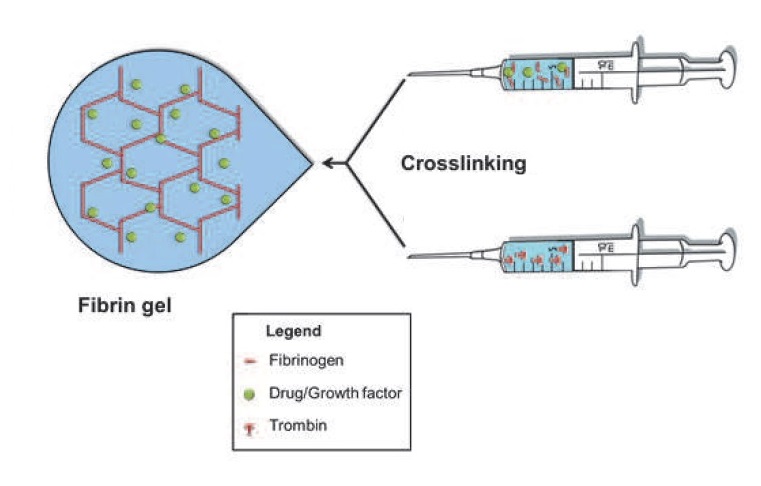}
\caption{The simplest form of gluing: fibrin gel preparation and drug/growth factor entrapment. Picture adapted from \cite{Fib}
}.
\label{fig:fibrin}
\end{figure}

\begin{figure}
\centering
\includegraphics[scale=0.4]{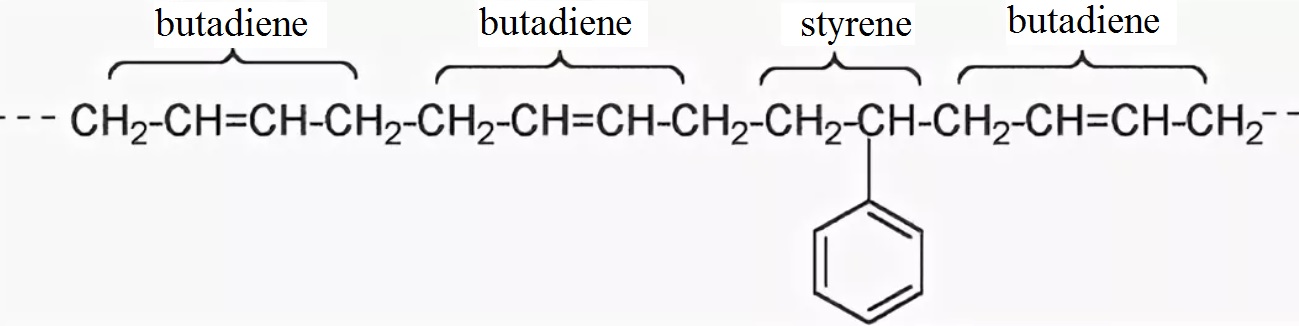}
\caption{Styrene - butadiene polymer
}.
\label{fig:buta}
\end{figure}

The synthesis of polymer chains with special bonding properties between them is an urgent task that finds numerous applications, because it determines the property of material made of such polymers . The gluing of polymers can have, for example, the form of photocrosslinking  - such polymers are used in biological applications (see \cite{Fib}); the scheme of this process in simplified form is represented at the picture \ref{fig:fibrin}. Random arrays of peptides is used in testing vaccines; the immunosignature effect withe them is considered in \cite{immun}. The task of synthesis biopolymers with anti bacterial properties is actual since the past century (from the plenty of recent results see, for example, \cite{r}). Assembly is often carried out from a pair of types of monomers - a typical example is styrene-butadiene rubber, the polymer of which is created from two types of monomers (see figure\ref{fig:buta}).

\section{Introduction}

However, the properties of polymer chain arrays depend on the type of their interaction, which in turn, is determined at the stage of their synthesis - as the subtle properties of the connection of monoblocks that determine their spatial location. In the simplest case, the interaction can be represented as the gluing of polymers in separate sections. In this article, we will consider the quantum control of the synthesis of abstract polymer chains from two monomeric links. This control, carried out through the entangled state of a pair of photons, allows us to achieve a degree of differentiation of the degree of gluing of different chains, which is unattainable with classical control.

The possibility of such a subtle distinction between the structure of polymer chains of the same nature is based on the property of quantum non locality of photons that control the synthesis of chains, namely, a violation of the Bell inequality for bi photons in the EPR state. 

The quantum non locality resulting from the experimentally confirmed violation of Bell's inequality is the most striking manifestation of the counter-intuitive nature of the microcosm. An experiment invented by Bell in the 60s of the twentieth century and implemented 20 years later (\cite{Zei},\cite{Asp}), showed the fundamental nature of quantum theory and the priority of its spirit over the classical understanding of nature. Discussions about the nature of quantum non locality have been going on for more than half a century (\cite{Kh}), approaches to understanding this nature are also offered through various information exchange protocols, for example, through non-local games (\cite{MA}). Photon entanglement experiments involve obtaining multidimensional entanglement from the angular momentum modes of bi-photon pairs (\cite{Zei2}, as well as four photons (\cite{Zei3}).

Non-locality formally follows from the existence of entangled multiparticle states, the simplest of which is the two-qubit state $\frac{1}{\sqrt 2}(|01\rangle\pm|10\rangle)$, where the $+$ sign gives a triplet $|t\rangle$, and the $-$ singlet $|s\rangle$. Both of these states have a maximum degree of entanglement, which is defined as entropy $ tr(\rho_1\ ln(\rho_1))$ the relative density matrix of the first qubit. Entanglement also underlies the so-called quantum acceleration of classical computing (\cite{Gr}). However, if the superiority of a scalable quantum computer requires the full-scale operation of an ensemble of hundreds of qubits, then on more limited models of computing this superiority can be demonstrated with a few qubits, which lies within the current capabilities of experimental physics. This paper is devoted to such a limited model-distributed computing with one-way control.

Finding out the role of quantum non locality in the management of distributed computing will allow us to better understand the nature of this phenomenon, but also promises to give a practical effect, allowing us to find more effective forms of such control. For bi-photon control, an example of such phenomenon is given in the paper \cite{Oz}. In this paper, we will strengthen this result by showing that bi-photonic control can also give the effect of differentiation between legal and nonlegal pairing of the chains assembled in the different locations. 

The criterion of quantum superiority over classical calculations in distributed computing is given not in terms of the calculation time, but in terms of the quality of the calculation result. The quality is detected after comparing the results in various remote nodes of the computing system, which requires communication between these nodes at light speed, so such protocols cannot exceed the light threshold for transmitting the information intended by the user. Nevertheless, it is precisely the non-locality that creates the qualitative superiority of the quantum model over classical distributed computing. It is interesting that the use of photonic quartets does not give advantage over the effect of bi-photons for this task. 

The differentiation of legal and nonlegal pairs of chains may play a role in the synthesis of bio polymers in living organism, namely for the immune response to the antigen.

In conclusion, we will give an interpretation of quantum non locality in the form of pre-programmed propagation of the wave function of a multiphoton system, which can intuitively clarify the mechanism of action of quantum non locality in distributed computing.

\section{Violation of Bell's inequality}

The experiment proposed by Bell (\cite{Be}) consists of the following. Two persons-Alice and Bob are given one qubit from a pair of qubits, after which each of them measures his qubit in a randomly selected one with probability $1/2$ the basis of two: for Alice, the choice is between the basis of the eigenvectors of the operators $\s_x$ and $\s_z$, for Bob - from the eigenvectors of the operators $\frac{1}{\sqrt 2}(\s_x+\s_z)$ and $\frac{1}{\sqrt 2}(\s_x-\s_z)$.
The result of the measurement is considered to be the eigenvalue corresponding to the measurement - in all cases it will be $\pm 1$. 

\begin{figure}
\begin{center}
\includegraphics[scale=0.6]{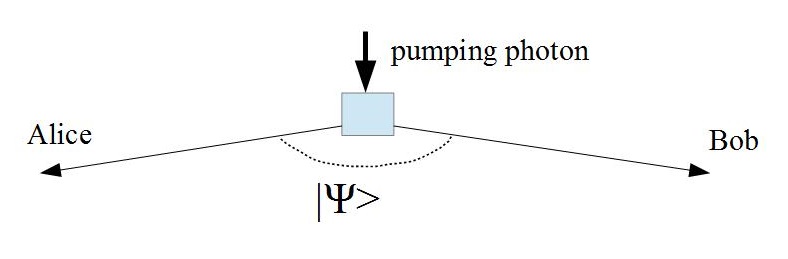}
\caption{Bell experiment. Alice chooses between $\s_x$ and $\s_z$ , Bob - between $\frac{1}{\sqrt 2}(\s_x+\s_z)$ and $\frac{1}{\sqrt 2}(\s_x-\s_z)$}.
\end{center}
\end{figure}

Denote by $a$ and $b$ results of Alice's measurement when selecting the basis corresponding to the operators $\s_x$ and $\s_z$ accordingly, and through $X$ and $Y$ - results of the Bean measurement when selecting the basis corresponding to the operators $\frac{1}{\sqrt 2}(\s_x+\s_z)$ and $\frac{1}{\sqrt 2}(\s_x-\s_z)$. After the measurement, we make a random variable $ \xi$ equal to $aX$, $aY$, $bX$ - when Alice and Bob choose the corresponding bases, and $ - bY$ - when choosing the bases corresponding to $b$ and $Y$. 

It is easy to see that if the measurement result of any participant does not depend on the orientation of the detector of the other participant, the mathematical expectation of the value $ \xi$ cannot be greater than $1/2$, since in this case we can write $M\xi=M\frac{1}{4} (aX+aY+bX-bY)=\frac{1}{4}M(a(X+Y)+b(X-Y))\leq 1/2$, where all occurrences of $a$, $b$, $X$, and $Y$ are the same numbers, and, consequently, one of the brackets is zero, and the other is modulo 2. 

However, suppose that the pair of photons distributed to Alice and Bob is in the $|t \rangle$ state. Then the quantum calculation of the mathematical expectation by the formula $M_\psi (A)=tr (A|\psi\rangle\langle\psi|) $ gives $M\xi = 1/\sqrt 2$. This means that the measurement result of one of the participants depends on the orientation of the detector of the other participant. This is quantum non locality. It proves the existence of instantaneous action at a distance, since the choice of the detector orientation can be made at the moment when the photon flies into Alice's laboratory, and nothing moving at light speed can reach Bob's laboratory to affect its result. 

"Knowing" about the orientation of Alice's detector "reaches" Bob's lab at a "speed" many times faster than light, which is contrary to the spirit of relativism. What is surprising is that there is no contradiction to the letter of the theory of relativity, since this channel cannot be used to transmit the information conceived by Alice to Bob at superluminal speed; to fix the violation of Bell's inequality, it is necessary to bring together the results of Alice's and Bob's measurements, which can only be done at the speed of light.

\section{The task of polymer chain synthesis}

The quantum advantage is based on entanglement, and the most explicit demonstration of it is given with the explicit use of non-locality. To this end, we will consider the task of distributed computing with one-way control. Here, the computing system consists of a mass of nodes distributed in space, whose operation is controlled by signals from a single central processor, which can send either classical states of bits - one for each node, or qubits that are in an entangled state, which we denote by $|\Psi\rangle$.

\begin{figure}
\begin{center}
\includegraphics[scale=0.4]{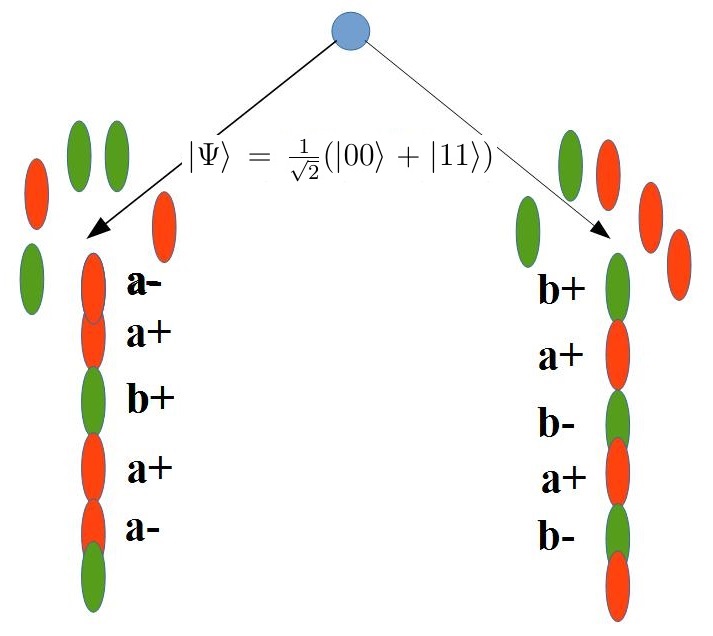}
 \caption{Simultaneous synthesis of a polymer pair. The sign is determined by measuring the photon for Alice (index 1) and Bob (index 2) in the observables:
}
$$
\begin{array}{lll}
&a_1=\sigma_x,\ &b_1=\sigma_z,\\
&a_2=\frac{1}{\sqrt 2}(\sigma_z-\sigma_x),\ &b_2=-\frac{1}{\sqrt 2}(\sigma_x+\sigma_z)
\end{array}
$$
\end{center}
\end{figure}
\hspace*{50mm}

In our task, we will talk about the simultaneous synthesis of abstract polymer chains at 4 points far from each other. The synthesis consists in joining a new link to the final link of the chain - a monomer taken at random from the reservoir associated with this chain. In this case, the connection of a new link is regulated by the result of measuring a photon that enters the detector connected to this chain. The measurement result is one of the numbers $ \pm 1$, and this number determines how to connect a new link to the chain: $+1$ means a connection with a forward shift, $-1$ - with a backward shift.

The result of the synthesis of 4 chains is determined by their overlapping each other in the form of an index of quality bonding: applying the better, the higher the index. Monomers - separate chain links, will be of two types, denoted by $a$ and $b$. Their structure is schematically shown in the figure.

The chain overlap index $Cr$ is the number of bonding monomers. Each monoblock has an outer (convex) and inner (concave) surface, where the latter is provided with a special ball located in its center. In a fixed position, two monoblock can be glued together in one of the following cases: 1) their surfaces or half of the surfaces are completely aligned by vertical displacement, or 2) their central balls are at the same point when this shift occurs, as shown in the figure \ ref{fig:polimers}.

The physical structure of the polymer on which the bonding depends is determined not only by the sequence of monoblocks in the chain; the bonding also depends on an additional option: their exact location relative to each other in the chain. Adjacent monoblocks in the polymer are connected by a flexible bond that can either shrink by $dx$, which is a quarter of the length of the monobloc, or stretch by the same length. In these cases, we will say that the monoblock is shifted back or forward, respectively, relative to the equilibrium position of the connection. During the synthesis, the monoblocks are set with these restrictions and their positions are fixed. Then the two chains are superimposed on each other and for each pair of overlapping monoblocks, the presence of gluing is established. From the accepted restriction, it follows that if in such a pair of overlapping monoblocks they were shifted in one direction, they are glued together in the same way as if there were no shifts; and if in different directions, the resulting shift is half the length of the monoblock.

Chain synthesis occurs as a sequential attachment to each of the existing chains of a new monoblock - the one that first appeared at the assembly point of one and the other chains. Monoblocks are taken from the environment surrounding the growth points, where they are in chaotic motion and both types are equally distributed. In this case, you can move the newly attached monoblock either backward or forward by a distance of $ dx $. We will denote the forward shift by $ + $, and the backward shift by $ - $. Each $j$ - th pair of monoblocks in both chains, superimposed on each other after synthesis, correspond, thus, to the four $ c_j^1c_j^2s_j^1s_j^2$, where the last two terms are shifts of $ s_j^{1,2} \in \{ +, - \} $.

\begin{figure}
\centering
\includegraphics[scale=0.6]{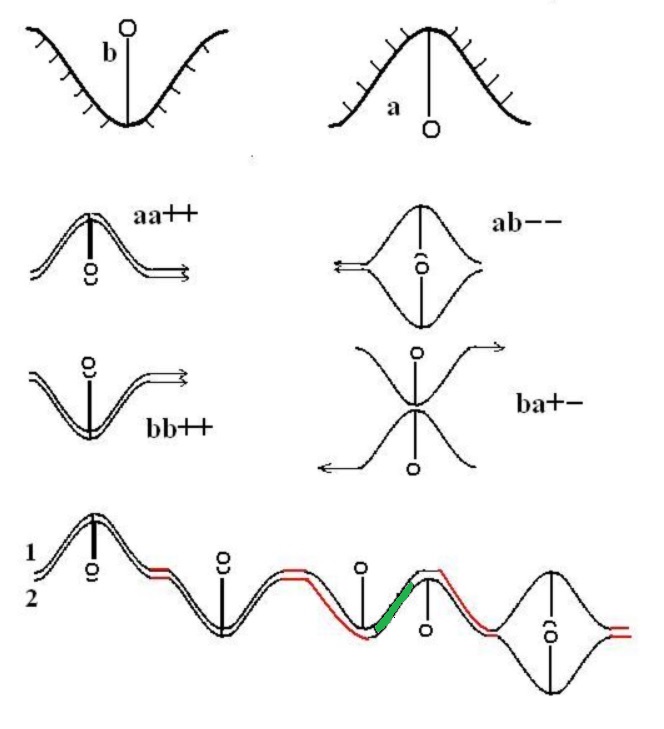}
\caption{Superposition of two polymers. The arrows indicate the direction of bond stretching (red) between adjacent monoblocks during polymer synthesis. Overlays of the form $aa++(--), ab++ (--), bb++ (--), ba+-(-+)$ give gluing, the rest of the gluing does not give. At the bottom, all pairs give glues.}
\label{fig:polimers}
\end{figure}

From our rules (see Figure \ref{fig:polimers}), it follows that the gluing corresponds to pairs of superimposed monoblocks of the form: $aa++ (--),\ ab++(--), bb++(--),ba+ - ( -+)$, while pairs of a different type:
$aa+-(-+), ab+ - (-+), bb+ - (-+), ab++ (--)$ do not give gluing. Note the asymmetric behavior of monoblocks of the type $a$ and $b$: the pairs $ab$ and $ba$ are glued together differently at the same shifts. This asymmetry looks like an asymmetry in the Bell inequality, which will give us an improvement in the quality of the resulting gluing under bi-photon control compared to classical control.

We assume that the growth of the polymer $ C_1 $ occurs at one point, and the growth of $ C_2 $ occurs at another, and these points are separated by a large distance (for example, they occur in different countries). The task is to organize this synthesis so that the number of non-glued pairs of superimposed monoblocks is minimal, or, in other words, that the number of glues is maximum.

In the paper \cite{Oz}, it is shown that the control of chain synthesis using bi-photon states of the form $|t \rangle$ gives a gain in the overlap index of more than $ 12\%$ compared to the classical control. This is shown by reducing it to Bell's inequality. The overlap index of two polymers can be calculated using the formula $Cr=\frac{1}{2}(1+NonCr)$, where $ NonCr$ is the number of overlays with good bonding. And the random variable $NonCr$, by virtue of our definitions, coincides with $ \xi$, so that its mathematical expectation is $1/\sqrt 2$, which gives the desired result.

\section{Case of three participants}
We assume the order of qubits: Alice, Natalia, Bob, and let the pair Alice, Bob be legitimate pair.
Maximum the differentiation of gluing indices for the three participants is 0.35 (0.5 for an illegitimate pair (which is the minimum possible value) and 0.85 for a legitimate pair (which is the maximum possible value)), where the legitimate pair has photons entangled with each other, and the illegitimate one does not. This value is achieved for the following conditions:

$\frac{1}{2}(|000\rangle+|010\rangle+|101\rangle+|111\rangle)=\frac{1}{2}(|0*0\rangle+1*1\rangle)(|*0*\rangle+|*1*\rangle)$

$\frac{1}{2}(|001\rangle+|010\rangle+|101\rangle+|110\rangle)$

$\frac{1}{\sqrt{2}}(|000\rangle+|011\rangle)$

$\frac{1}{\sqrt{2}}(|001\rangle+|010\rangle)$

$\frac{1}{\sqrt{2}}(|010\rangle+|111\rangle)$

$\frac{1}{\sqrt{2}}(|011\rangle+|110\rangle)$

For the case of four syllables, it is possible to recognize a legitimate couple - knowing that the 3rd participant is included in it always, and the state itself is received as tensor product of 2 states entangled pair and two states an illegitimate participant, you can see which positions (1 or 2) in the final state is 1 and 0 so, in order to "bind" with the 3 state to blow or two combinations 00 and two 11 or two combinations 01 and two 10. For the case of three syllables, a similar method is suitable.

The maximum level of gluing for a pair is achieved only under the above states of a legitimate pair. Thus, it is possible to obtain all the listed states by enumerating various values of the state of the not entangled with others participant.

the maximum sum E(NonCr) for all opposite-sex couples of three people is 1.157, which is less than the same value for classical management (0.75 * 2).

\section{Photon quartets for 4-chain synthesis}

We generalize the result of \cite{Oz} to 4 polymer chains divided into pairs by gender. Let's say we have 4 participants: Alice, Natasha, Bob, and Ivan, and each of them synthesizes their polymer chain according to the same rules as in the previous task, so that Alice and Natasha choose the observables between $\s_x$ and $\s_z$, and Bob and Ivan choose between $\frac{1}{\sqrt 2}(\s_x+\s_z)$ and $\frac{1}{\sqrt 2}(\s_x - \s_z)$ respectively, regardless of each other. If you divide the participants into fixed pairs, for example, Alice-Bob, Natasha-Ivan, and give each pair a bifoton, as in the previous paragraph, the winnings will be the same as before. But if this control calculates the index of gluing an alternative pair, for example, Natasha and Bob, we will get a lower index then in the classical control, since there is no coherence between different bi-photons.

Conditions with maximum gluing differentiation "legitimate - illegitimate" couples:
There are 4 states that give a good overlap for Natalia-Bob and Alice-Ivan pairs (0.85) and for Natalia-Ivan, Alice-Bob pairs, bad ( 0.5): these are states

$\frac{1}{2}(|0011\rangle+|0110\rangle+|1001\rangle+|1100\rangle)$

$\frac{1}{2}(|0010\rangle+|0111\rangle+|1000\rangle+|1101\rangle)$

$\frac{1}{2}(|0001\rangle+|0100\rangle+|1011\rangle+|1110\rangle)$

$\frac{1}{2}(|0000\rangle+|0101\rangle+|1010\rangle+|1111\rangle)$

There are also 4 states that give bad overlap for Natalia-Bob and Alice-Ivan pairs and good overlap for Natalia-Ivan and Alice-Bob pairs:

$\frac{1}{2}(|0010\rangle+|0100\rangle+|1011\rangle+|1101\rangle)$

$\frac{1}{2}(|0001\rangle+|0111\rangle+|1000\rangle+|1110\rangle)$

$\frac{1}{2}(|0000\rangle+|0110\rangle+|1001\rangle+|1111\rangle)$

$\frac{1}{2}(|0011\rangle+|0101\rangle+|1010\rangle+|1100\rangle)$

Also, the maximum difference in the values of the gluing indices is achieved only when entangled bi-photons are distributed to legitimate pairs.

The maximum sum E(NonCr) for all 4 opposite-sex couples will be 2.7071067811865475, which means that for each pair it is impossible to win over the classical control (the sum should have been> 0.75 * 4 = 3).

\begin{figure}
\begin{center}
\includegraphics[scale=0.6]{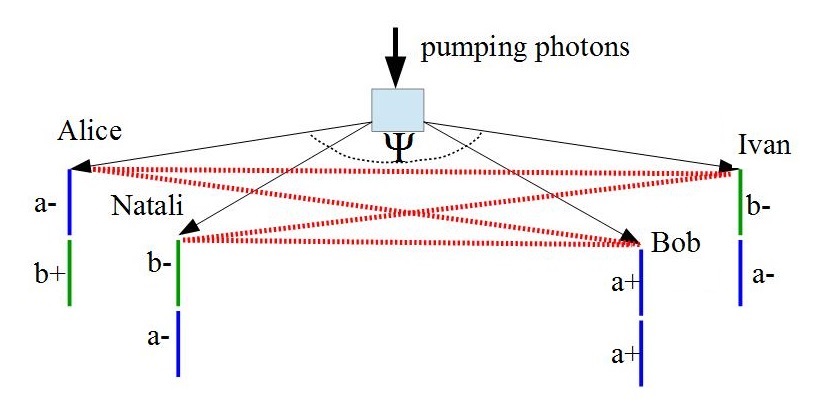}
\caption{Synthesis of 4 chains. The dotted line shows all heterosexual pairs of polymers, the superposition of which at $|\Psi\rangle\in{(1)}$ has a higher quality than when using bi-photon control.}.
\end{center}
\end{figure}

\section{Minimal overlap}
Also, for 3 and 4 participants, there are states where E(NonCr) for each heterosexual pair becomes the minimum possible - 0.5 (including less than with classical control)

$\frac{1}{2}(|000\rangle+|001\rangle+|110\rangle+|111\rangle)$

$\frac{1}{2}(|010\rangle+|011\rangle+|100\rangle+|101\rangle)$

$\frac{1}{2}(|0000\rangle+|0011\rangle+|1100\rangle+|1111\rangle)$

$\frac{1}{2}(|0100\rangle+|0111\rangle+|1000\rangle+|1011\rangle)$

For example, consider the state $\frac{1}{2}(|0000\rangle+|0011\rangle+|1100\rangle+|1111\rangle)$ for Alice, Nataly, Bob and Ivan. Let's build a reduced density matrix for each heterosexual pair. We will look at the values at the corresponding places of the density matrix (we restrict ourselves only to those values for which at least one in a pair has a nonzero amplitude value in the initial state)

Therefore, for any pair and any choice of observables (A), it will be true that $tr(A\rho)$ = 0. Therefore, E(NonCr) = $\frac{1}{2}$(1+0) = $\frac{1}{2}$ for all opposite-sex couples.

\section{Conclusion}

We have consider the task of the comparison on gluing index in all heterosexual pairs of chains obtained by simultaneous synthesis of 3 or 4 polymer chains at the remote locations under one-way control. We found that the result in the different formulations strongly depends on the controlling signals. The use of EPR biphoton pairs allow to reach the maximal difference of gluing degree (kinship) between legal and nonlegal pairs and the minimal gluing degree (antagonism) in all heterosexual pairs - both results are unattainable for the classical control. 

If the aim is to increase the kinship of legal pair and at the same time to increase the antagonism for the illegal pair, then we must distribute the EPR pair between the legal participants and keep illegal one disentangled with them. The difference in the index of gluing between the legal and non legal pairs will be then $0.35$.  

If the aim is to reach the minimal possible index of gluing in all heterosexual pair (total antagonism) we must distribute bi-photonic signals to homosexual pairs and keep the different sexes non entangled. 

This non symmetric behavior between the ways of success in the formally opposite tasks can be expressed shortly. To get the maximal antagonism between sexes we must entangle pairs with the same sex; to get the maximal difference between legal and nonlegal heterosexual pair we must entangle the legal heterosexual pair exclusively. 

This conclusion is made only for our artificially build example. But this example directly rests on the main feature of entanglement - the violation of Bell inequality, hence the conclusion applies to the entangled control for the more wide class of chains assembled in the remote locations that may have interest for quantum biology.

\section{ Acknowledgements }

The paper was prepared in the Moscow Center for Fundamental and Applied Mathematics.


\begin{thebibliography}{99}
\bibitem{Fib}P. Ferreira, J. F. J. Coelho, J. F. Almeida and M. H. Gil (2011). Photocrosslinkable Polymers for Biomedical Applications, Biomedical Engineering - Frontiers and Challenges, Prof. Reza Fazel (Ed.), ISBN: 978-953-307-309-5, InTech, Available from: http://www.intechopen.com/books/biomedical-engineering-frontiers-andchallenges/photocrosslinkable-polymers-for-biomedical-applications
\bibitem{immun}Stafford, Phillip; Halperin, Rebecca; Legutki, Joseph Bart; Magee, Dewey Mitchell; Galgiani, John; Johnston, Stephen Albert (2012-04-01). "Physical Characterization of the "Immunosignaturing Effect"". Molecular \& Cellular Proteomics. 11 (4): M111.011593. doi:10.1074/mcp.M111.011593.
\bibitem{r} C. RAJAM AND D. ROOP SINGH,  SYNTHESIS OF CERTAIN RANDOM COPOLYMERS CONTAINING ARYLIDENE DIOL MOIETY AND STUDIES ON THEIR ANTIBACTERIAL ACTIVITY AND PHOTOCROSSLINKING EFFICACY OF BLEND NANOFIBERS, International Journal of Science, Engineering and Technology Research (IJSETR), Volume 4, Issue 1, January 2015.
\bibitem{Zei}Jian-Wei Pan; D. Bouwmeester; M. Daniell; H. Weinfurter; A. Zeilinger (2000). "Experimental test of quantum non locality in three-photon GHZ entanglement". Nature. 403 (6769): 515–519.
\bibitem{Asp}Aspect, Alain; Dalibard, Jean; Roger, Gérard (December 1982). "Experimental Test of Bell's Inequalities Using Time- Varying Analyzers". Physical Review Letters. 49 (25): 1804–1807.
\bibitem{Kh}A.Khrennikov, After Bell, Fortschritte der Physik (Progress in Physics) 65, N 6-8, 1600014 (2017).
\bibitem{MA}Mateus Araújo, Flavien Hirsch, Marco Túlio Quintino, Quantum 4, 353 (2020)
DOI:10.22331/q-2020-10-28-353.
\bibitem{Zei2}Jaroslav Kysela, Manuel Erhard, Armin Hochrainer, Mario Krenn, Anton Zeilinger, Experimental High-Dimensional Entanglement by Path Identity, PNAS 117(42), 26118-26122 (2020)
DOI:10.1073/pnas.2011405117.
\bibitem{Zei3}Manuel Erhard, Mehul Malik, Mario Krenn, Anton Zeilinger, Experimental GHZ Entanglement beyond Qubits, Nature Photonics 12, 759-764 (2018)
DOI:10.1038/s41566-018-0257-6.
\bibitem{Gr}L.Grover, A fast quantum mechanical algorithm for database search, Proceedings, 28th Annual ACM Symposium on the Theory of Computing (STOC), May 1996, pages 212-219. Proceedings, Melville, NY, 2006, vol. 810.
\bibitem{Oz}Y.Ozhigov, Distributed synthesis of chains quth one-way bi-photonic control, Quantum Information and Computation, Vol. 18, No. 7\&8 (2018) 0592-0598.
\bibitem{Be}J. Bell,  "On the Einstein Podolsky Rosen Paradox"; Physics, (1964), 1 (3): 195–200.
%\bibitem{Fe}R.Feynman, QED: The strange theory of linght and matter, Princeton University Press, 1985.
%\bibitem{FeH}R.Feynman, D.Hibbs, Quantum mechanics and path integrals, 1984,
%R. Feynman, D. Hibbs, Quantum Mechanics in Path integrals, Moscow, Nauka, Phys. - mat. lit. 
%\bibitem{Xi}G. Xiuhong, S. Albeverio, C. Kai, F. Shaoming, L. Xianqing, Entanglement of formation and concurrence for mixed states, Front. Comput. Sci. China 2008, 2(2): 114–128DOI 10.1007/s11704-008-0017-8
\end{thebibliography}
\end{document}